\documentclass[aps,pra,twocolumn,groupedaddress]{revtex4-1}
\usepackage{color}
\usepackage{amsmath,amsfonts,amssymb}
\usepackage{amsthm}
\usepackage{bm}
\usepackage{graphicx}
\usepackage{subfigure}
\usepackage{bbm}
\usepackage{epstopdf}
\usepackage{epsfig}
\usepackage{verbatim}
\usepackage{array}
\usepackage{ulem}
\usepackage{notes2bib}
\usepackage[T1]{fontenc}
\usepackage{tgtermes}

\usepackage{dcolumn}
\usepackage{physics}
\usepackage{braket}
\usepackage[unicode=true,bookmarks=true,bookmarksnumbered=false,bookmarksopen=false, breaklinks=true,pdfborder={0 0 1},backref=false,colorlinks=true]{hyperref}
\hypersetup{linkcolor=magenta, urlcolor=blue, citecolor=blue, pdfstartview={FitH}, unicode=true}

\begin{document}

\title{Fermionic superfluid transition and pairing scale separation in the kagome flat band}

\author{Xiang Li}
\affiliation{School of Physics and Technology, Wuhan University, Wuhan 430072, China}
\author{Yu Wang}
\email{yu.wang@whu.edu.cn}
\affiliation{School of Physics and Technology, Wuhan University, Wuhan 430072, China}

\date{\today}

\begin{abstract}
Flat-band systems provide a promising platform for exploring exotic correlated phases owing to quenched kinetic energy and enhanced interaction effects. In this work, we investigate the flat-band superfluid transition in the attractive kagome Hubbard model using large-scale determinant quantum Monte Carlo simulations accelerated by fast Fourier transforms and delay updates. Through systematic calculations of the transition temperature, spin susceptibility, and density compressibility, we establish the finite-temperature phase diagram at the flat-band filling $\rho=5/3$ and find a clear linear scaling of the superfluid transition temperature with the interaction strength, $T_{\mathrm{c}}=0.0314|U|$.
Our results reveal a wide separation of temperature scales between local pair formation and global phase coherence, which stabilizes a broad intermediate supercompressible bosonic fluid region of preformed pairs that bridges the normal Fermi gas and the flat-band superfluid. We further demonstrate that both the transition temperature and density compressibility are significantly enhanced in the flat band compared to van Hove fillings. Our findings highlight the essential role of flat-band-enhanced pair-density fluctuations in separating the pairing and phase-coherence scales, offering clear experimental signatures for probing flat-band superfluidity and superconductivity in cold-atom and kagome-material platforms.
\end{abstract}

\maketitle
\textit{Introduction.} Flat-band systems provide a compelling platform for exploring strongly correlated quantum phases, representing a paradigm shift where the quenching of kinetic energy drives robust superfluidity and superconductivity~\cite{QG1,basicreview1,basicreview2,tbgtc1,tbgtc2,tbgtc3,tbgtc4,tbgtc5,linearTctbg1,topoflat1,topoflat2,topoflat3,compositekag3,compositeLieb1,compositeLieb2,compositeLieb3,compositekag2}. In a dispersionless band,  characterized by a vanishing single-particle group velocity and a  diverging effective mass, electrons are conventionally expected to localize into an insulating state. However, the extreme suppression of kinetic energy allows incredibly weak electron-electron interactions to dominate, enhancing interaction-driven Cooper pairing~\cite{tbgtc1,tbgtc2,tbgtc3}. In a system of essentially immobile fermions, the macroscopic phase coherence and dissipationless transport of Cooper pairs are governed by the nontrivial quantum geometry of the underlying Bloch wavefunctions~\cite{QG1}. Specifically, the quantum metric, which quantifies the distance between adjacent quantum states in momentum space, generates a finite superfluid weight~\cite{QG1}. This geometric contribution enables the macroscopic propagation of bound pairs despite the absence of single-particle mobility. The observation of flat-band superconductivity in magic-angle twisted bilayer graphene  establishes the experimental reality of this quantum-geometric framework~\cite{tbgtc1,tbgtc2,tbgtc3,tbgtc4,tbgtc5}.

Geometric flat-band superconductivity features a transition temperature $T_{\mathrm{c}}$ that scales linearly with the interaction strength ($T_{\mathrm{c}} \propto |U|$) rather than exponentially ($T_{\mathrm{c}} \propto \exp[-1/|U|]$) as in conventional BCS theory~\cite{QG1,linearTctbg1,topoflat1,topoflat2,topoflat3,compositekag3,compositeLieb1,compositeLieb2}. This linear dependence implies that even modest attractive interactions can induce extraordinarily high transition temperatures relative to the underlying energy scales, establishing quantum-geometrically nontrivial flat bands as a promising route toward high-temperature superfluidity and superconductivity. In recent years, exploring these flat-band systems has become a frontier in the design and discovery of high-temperature superconductors~\cite{QG1,topoflat1,topoflat2,topoflat3,compositekag3,compositeLieb1,compositeLieb2,compositeLieb3}. 

In magic-angle moir\'e superlattices, the interlayer tunneling quenches the Fermi velocity, flattening the minibands~\cite{twistreview1,twistreview2,twistreview3,twistreview4}. Theoretical studies of the superconducting critical temperature $T_{\mathrm{c}}$ in moir\'e flat bands, particularly those in twisted bilayer graphene, estimate $k_{\mathrm{B}}T_{\mathrm{c}}\sim10^{-4}|U|$~\cite{linearTctbg1}, consistent with experimental observations typically ranging from $T_{\mathrm{c}} \approx 1.0$ to $3.0$ K~\cite{tbgtc1,tbgtc2,tbgtc3,tbgtc4,tbgtc5}. 

Topological flat-band models, where band flatness is achieved by engineering the hopping integrals, are of considerable theoretical interest because they provide a controllable setting for tailoring quantum geometry and designing high-temperature superconductors~\cite{QG1,topoflat1,topoflat2,topoflat3}. In these systems, determinant quantum Monte Carlo (DQMC) studies predict a linear $T_{\mathrm{c}}$ scaling $k_{\mathrm{B}}T_{\mathrm{c}} \sim 10^{-2}|U|$~\cite{topoflat2,topoflat3}, suggesting that such flat bands are promising candidates for high-temperature superconductivity. However, realizing superfluidity with ultracold fermions in these topological flat-band models poses significant experimental challenges, as engineering the required hopping parameters in optical lattices relies on sophisticated techniques such as Floquet driving~\cite{synthetic1,synthetic2} or laser-assisted tunneling~\cite{synthetic3}. 

In kagome and Lieb lattices, the emergence of a dispersionless band is driven not by external tuning, but by the destructive quantum interference of  single-particle hopping paths, an effect dictated entirely by the underlying lattice geometry~\cite{compositekag2,compositekag3,compositeLieb1,compositeLieb2,compositeLieb3}. In these composite lattices, mean-field calculations within the attractive Hubbard model predict a benchmark transition temperature $T_{\mathrm{c}}=0.1t$ (where $t$ is the hopping amplitude) for flat-band superconductivity and superfluidity~\cite{compositekag3,compositeLieb2}. For a solid-state realization, a hopping scale of a few tens of meV would put this temperature on the order of tens of kelvin.

In moir\'e systems, the relatively low $T_{\mathrm{c}}$ is inherently limited by the miniaturized moir\'e Brillouin zone, which drastically restricts the carrier density available for pairing. By contrast, engineered topological flat-band models, kagome and Lieb systems can support significantly higher transition temperatures, as the pairing states span the full lattice Brillouin zone.  
Compared to other flat-band systems, the inherent geometric frustration of the kagome lattice suppresses competing crystalline orders~\cite{FFTdelay2026}, making it a clean platform for exploring flat-band superfluidity. In particular, the intrinsic kagome flat band yields a mean-field benchmark of the transition temperature $T_{\mathrm{c}} = 0.1t$~\cite{compositekag3}, exceeding the values predicted for engineered topological flat-band models. For ultracold gases, this $T_{\mathrm{c}}$ aligns with the experimentally relevant superexchange temperature regime. Consequently, the attractive kagome Hubbard model offers a realistic route toward flat-band superfluidity, particularly given that recent cold-atom experiments have realized optical kagome lattices~\cite{kagomefbexp1} and demonstrated the ability to populate the upper flat band~\cite{kagomefbexp2}. While mean-field approaches neglect fluctuations and typically overestimate the transition temperature, unbiased DQMC studies of the kagome flat-band superfluid transition remain rare, as computational bottlenecks  historically restricted accessible system sizes and precluded definitive finite-size scaling analyses ~\cite{QMCkagome0,QMCkagome1,QMCkagome2,QMCkagome3,QMCkagome4,QMCkagome5,QMCkagome6}. 

Recently, the implementation of fast Fourier transforms (FFT) and delay updates has effectively extended the reach of DQMC to large-scale composite lattices~\cite{FFTdelay2026}. Using this accelerated DQMC approach, we investigate the flat-band superfluid transition in the attractive kagome Hubbard model. Our simulations demonstrate that the system enters the flat-band superfluid state via a high-temperature supercompressible bosonic fluid (SCBF) phase, occurring at temperature and interaction scales highly relevant to current cold-atom experiments.

\label{sec:model}
\textit{Model and method.} We consider the attractive Hubbard model on the kagome lattice, described by the Hamiltonian
\begin{figure}
  \includegraphics[width=0.75\linewidth]{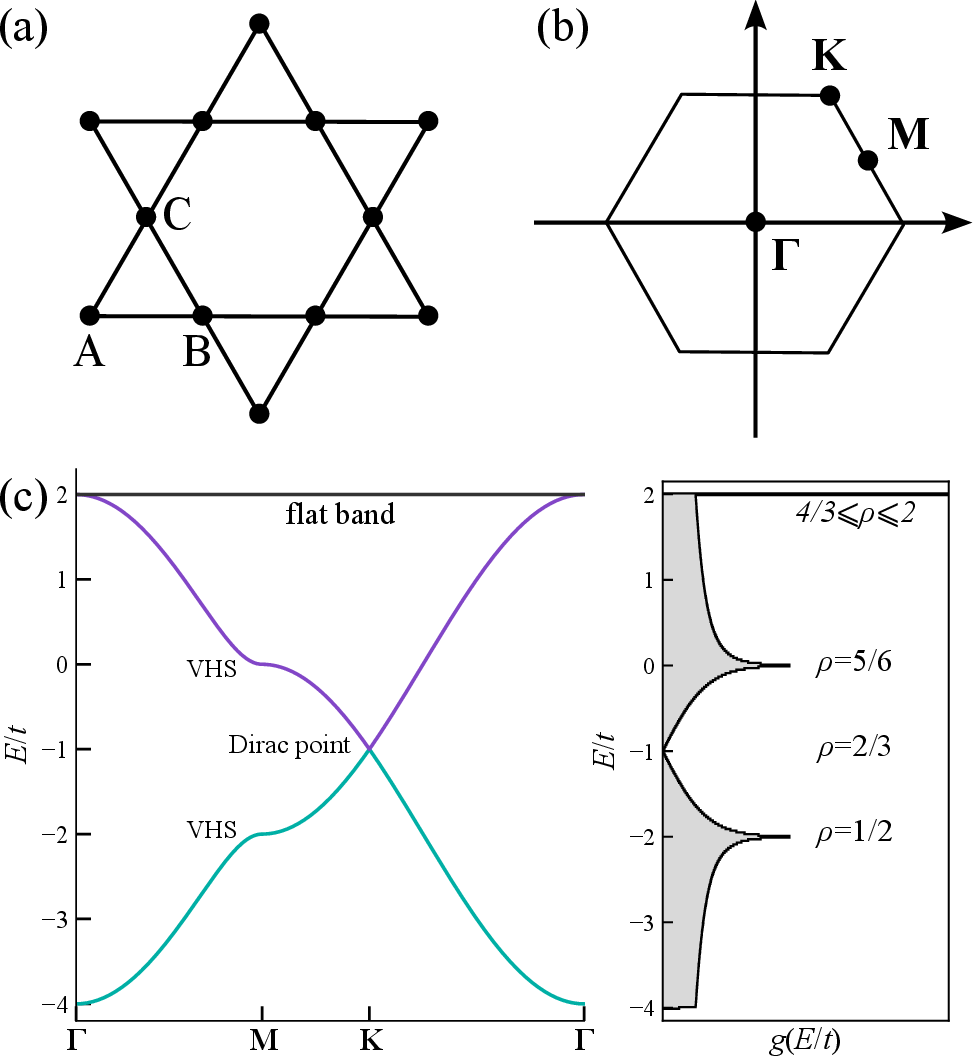}%
  \caption{\label{fig:band}(a) Structure of the kagome lattice, consisting of a triangular Bravais lattice with three distinct sublattices A, B, and C. (b) The corresponding first Brillouin zone, with the high-symmetry points $\boldsymbol{\Gamma}$, $\mathbf{K}$, and $\mathbf{M}$ labeled. (c) Band structure along the high-symmetry path $\boldsymbol{\Gamma}$-$\mathbf{M}$-$\mathbf{K}$-$\boldsymbol{\Gamma}$ (left panel) and the corresponding density of states (right panel). The van Hove singularities (VHS) occur at fillings $\rho=5/6, 1/2$, while the Dirac points correspond to $\rho=2/3$. Flat-band occupation spans the filling range $4/3 \leqslant \rho \leqslant 2$. 
  }
\end{figure}
\begin{equation}
\label{eq:model}
H=-t\sum_{\langle ij\rangle,\sigma}\left(c^\dagger_{i\sigma}c_{j\sigma}+\mathrm{H.c.}\right)
+U\sum_i n_{i\uparrow}n_{i\downarrow}
-\mu\sum_{i,\sigma}n_{i\sigma},
\end{equation}
where $\langle ij\rangle$ denotes pairs of nearest-neighbor sites. The operator $c^\dagger_{i\sigma}$ ($c_{i\sigma}$) creates (annihilates) a fermion with spin $\sigma=\uparrow,\downarrow$ at site $i$, and $n_{i\sigma}=c^\dagger_{i\sigma}c_{i\sigma}$ is the particle-number operator. Hereafter, we set $k_{\mathrm{B}}=1$ and take the hopping amplitude $t$ as the unit of energy. The parameter $U<0$ denotes the attractive Hubbard interaction, and the chemical potential $\mu$ tunes the average particle filling $\rho$.

The kagome lattice is composed of a triangular Bravais lattice with a three-site basis, as illustrated in Fig.~\ref{fig:band}(a). Its first Brillouin zone and high-symmetry points are shown in Fig.~\ref{fig:band}(b). In momentum space, the kinetic term reads
\begin{equation}
H_0=\sum_{\mathbf{k},\sigma}
\Psi_{\mathbf{k}\sigma}^{\dagger}
\mathcal{H}_0(\mathbf{k})
\Psi_{\mathbf{k}\sigma},
\end{equation}
where $\Psi_{\mathbf{k}\sigma}=(c_{\mathbf{k}\mathrm{A}\sigma},c_{\mathbf{k}\mathrm{B}\sigma},c_{\mathbf{k}\mathrm{C}\sigma})^T$ is the fermion spinor defined in the three-sublattice basis of the kagome unit cell, and
\begin{equation}
\label{eq:H0}
\mathcal{H}_0(\mathbf{k})=-2t
\begin{pmatrix}
0 & \cos k_1 & \cos k_2\\
\cos k_1 & 0 & \cos k_3\\
\cos k_2 & \cos k_3 & 0
\end{pmatrix}.
\end{equation}
Here, $k_n=\mathbf{k}\cdot\boldsymbol{\delta}_n$, with the vectors $\boldsymbol{\delta}_1=(1,0)$, $\boldsymbol{\delta}_2=(-1,\sqrt{3})/2$, and $\boldsymbol{\delta}_3=(-1,-\sqrt{3})/2$. Diagonalizing Eq.~(\ref{eq:H0}) yields two dispersive bands, $\varepsilon_{\pm}(\mathbf{k})=t\left[-1\pm\sqrt{4\sum_{n=1}^{3}\cos^2 k_n-3}\right]$, alongside an exactly flat top band, $\varepsilon_0(\mathbf{k})=2t$. Figure~\ref{fig:band}(c) shows these bands and the corresponding density of states. The dispersive bands exhibit van Hove singularities at the $\mathbf{M}$ point at fillings $\rho=5/6, 1/2$, and feature Dirac crossings at the $\mathbf{K}$ point at $\rho=2/3$. The flat band is populated within the filling range $4/3 \leqslant \rho \leqslant 2$.
\begin{figure}[t]
  \centering
  \includegraphics[width=0.75\linewidth]{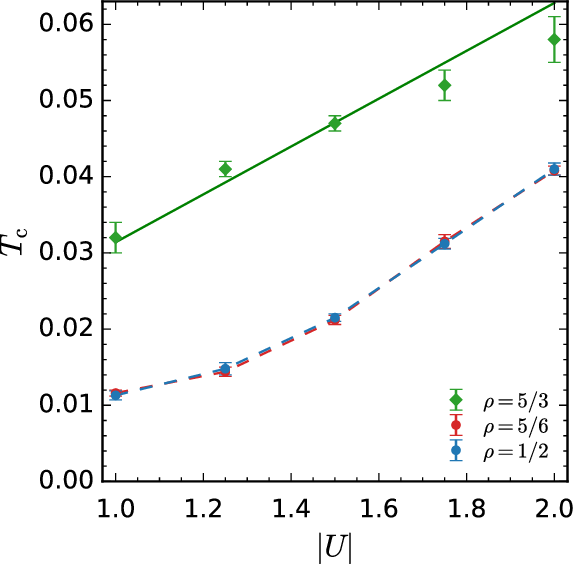}
  \caption{\label{fig:Tc}Dependence of the superfluid transition temperature $T_{\mathrm{c}}$ on the attractive interaction strength $|U|$ at the flat-band filling $\rho=5/3$ and the two van Hove fillings $\rho=5/6, 1/2$. The solid green line denotes a linear fit given by $T_{\mathrm{c}}=0.0314|U|$, and the dashed lines serve as guides to the eye.}
\end{figure}

We combine FFT and delay-update techniques in our DQMC simulations to study the attractive Hubbard model on the kagome lattice with linear sizes ranging from $L=8$ to $18$. Specifically, we simulate the superfluid transition temperature, spin susceptibility and density compressibility across the weak-to-intermediate coupling regime at the flat-band filling $\rho=5/3$. For comparison, we also calculate the transition temperature and density compressibility at the two van Hove fillings.

\label{sec:SFTc}
\textit{Transition temperature.} We extract the superfluid transition temperature $T_{\mathrm{c}}$ using a finite-size scaling analysis of the dimensionless correlation ratio~\cite{FFTdelay2026},
\begin{equation}
\label{eq:Rpair}
R_{\mathrm{Pair}}(T, L)=1-
\frac{S_{\mathrm{Pair}}(L,\boldsymbol{\Gamma}+\delta\mathbf{k})}
{S_{\mathrm{Pair}}(L,\boldsymbol{\Gamma})},
\end{equation}
where $S_{\mathrm{Pair}}(L,\mathbf{Q})=\frac{1}{N}\sum_{i,j}e^{i\mathbf{Q}\cdot(\mathbf{r}_i-\mathbf{r}_j)}\langle c_{i\uparrow}^{\dagger}c_{i\downarrow}^{\dagger}c_{j\downarrow}c_{j\uparrow}+c_{j\downarrow}c_{j\uparrow}c_{i\uparrow}^{\dagger}c_{i\downarrow}^{\dagger}\rangle$ is the pairing structure factor. Here, $N=3L^2$ is the total number of lattice sites for a system of linear dimension $L$, the point $\boldsymbol{\Gamma}$ denotes the Brillouin zone center, and $\delta\mathbf{k}$ is the smallest non-zero wavevector allowed by the finite lattice geometry. For large $L$, $R_{\mathrm{Pair}}(T, L)\rightarrow 1$ in the superfluid phase, whereas $R_{\mathrm{Pair}}(T, L)\rightarrow 0$ when the superfluid order vanishes. At the critical point, $R_{\mathrm{Pair}}(T_{\mathrm{c}}, L)$ is a universal constant independent of $L$. Therefore, the transition temperature $T_\mathrm{c}$ can be identified as the single temperature at which $R_{\mathrm{Pair}}(T, L)$ curves for different system sizes intersect. Following Refs.~\cite{cross1,cross2,cross3}, we extract $T_{\mathrm{c}}$ from the crossing point of the $R_{\mathrm{Pair}}(T, L)$ curves corresponding to the two largest system sizes. We perform this finite-size scaling analysis of $R_{\mathrm{Pair}}(T, L)$ in the coupling regime $1.0\leqslant |U|\leqslant 2.0$ and estimate the transition temperature $T_\mathrm{c}$ at the flat-band filling $\rho=5/3$ and the two van Hove fillings (see Supplemental Material~\cite{SupplementalMaterial} for details).
\begin{figure}[t]
  \centering
  \includegraphics[width=0.75\linewidth]{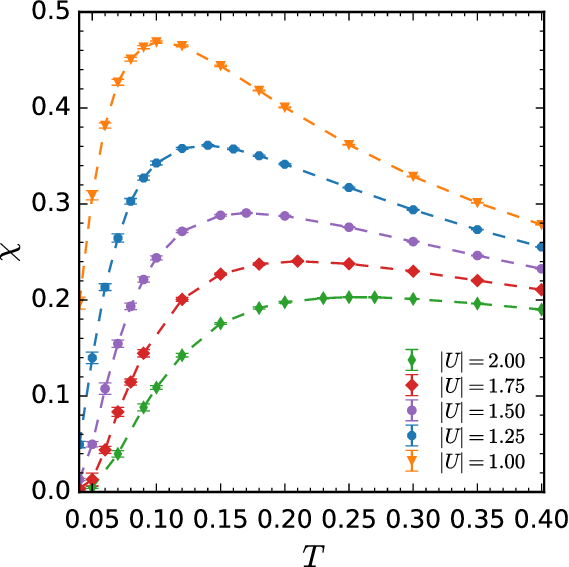}
  \caption{\label{fig:sus}Temperature dependence of the spin susceptibility $\chi$ for various attractive interaction strengths $|U|$ on an $L=18$ kagome lattice at the flat-band filling $\rho=5/3$. Dashed lines serve as guides to the eye.}
\end{figure}

Figure~\ref{fig:Tc} summarizes the dependence of the superfluid transition temperature $T_{\mathrm{c}}$ on the attractive interaction strength $|U|$ at the flat-band filling $\rho=5/3$ and the two van Hove fillings $\rho=5/6, 1/2$. In the weak-to-intermediate coupling regime $1.0\leqslant|U|\leqslant2.0$, $T_{\mathrm{c}}$ at the flat-band filling is substantially higher than the corresponding transition temperature at either van Hove filling. 
The flat-band $T_{\mathrm{c}}$ exhibits an approximately linear scaling with $|U|$, in agreement with previous studies~\cite{QG1,linearTctbg1,topoflat1,topoflat2,topoflat3,compositekag3,compositeLieb1,compositeLieb2}. A linear fit yields $T_{\mathrm{c}}=0.0314|U|$, revealing a steeper slope than those reported for topological flat-band models~\cite{topoflat2,topoflat3}. At $|U|=2.0$, the flat-band $T_{\mathrm{c}}$ reaches $0.06t$, which is comparable to the low temperatures achieved in recent cold-atom experiments ($T=0.05^{+0.06}_{-0.05}$)~\cite{Xutemperature2025}. The upward trend of $T_{\mathrm{c}}$ with $|U|$ in this regime suggests that even higher transition temperatures can be achieved at stronger attractions, though the coupling regime $|U| > 2$ lies beyond the reach of our DQMC simulations due to large statistical uncertainties (see Supplemental Material~\cite{SupplementalMaterial} for details).

\label{sec:spinsus}
\textit{Spin susceptibility.} The onset of spin-singlet pair formation is characterized by a temperature scale $T_{\mathrm{S}}$, which is identified by the maximum of the spin susceptibility, 
\begin{equation}
\label{eq:chi}
\chi=\frac{3\beta}{4N} \left[ \left\langle\left(\sum_i(n_{i\uparrow}-n_{i\downarrow})\right)^2\right\rangle -\left\langle\sum_i(n_{i\uparrow}-n_{i\downarrow})\right\rangle^2 \right].
\end{equation}
Above $T_{\mathrm{S}}$, cooling increases $\chi$ by enhancing the polarizability of unpaired spins, whereas below $T_{\mathrm{S}}$, cooling promotes pairing and consequently decreases $\chi$~\cite{topoflat2,topoflat3}. This susceptibility is readily accessible in cold-atom experiments through measurements of spin fluctuations~\cite{Schafer2020,Hartke2023}.

Figure~\ref{fig:sus} shows the temperature dependence of $\chi$, calculated for various attractive interaction strengths $|U|$ on an $L=18$ kagome lattice at the flat-band filling $\rho=5/3$.
The susceptibility exhibits a maximum that shifts systematically to higher temperatures with increasing $|U|$. We estimate $T_{\mathrm{S}} = 0.10, 0.14, 0.17, 0.21$, and $0.25$ for $|U|=1.00, 1.25, 1.50, 1.75$, and $2.00$, respectively. Across the interaction regime investigated, the pair-formation temperature $T_{\mathrm{S}}$ exceeds the superfluid transition temperature $T_{\mathrm{c}}$ by a factor of several. This wide separation of temperature scales confirms that preformed pairs emerge well before the onset of macroscopic phase coherence.

\label{sec:comp}
\textit{Density compressibility.}
\begin{figure}[t]
  \centering
  \includegraphics[width=1\linewidth]{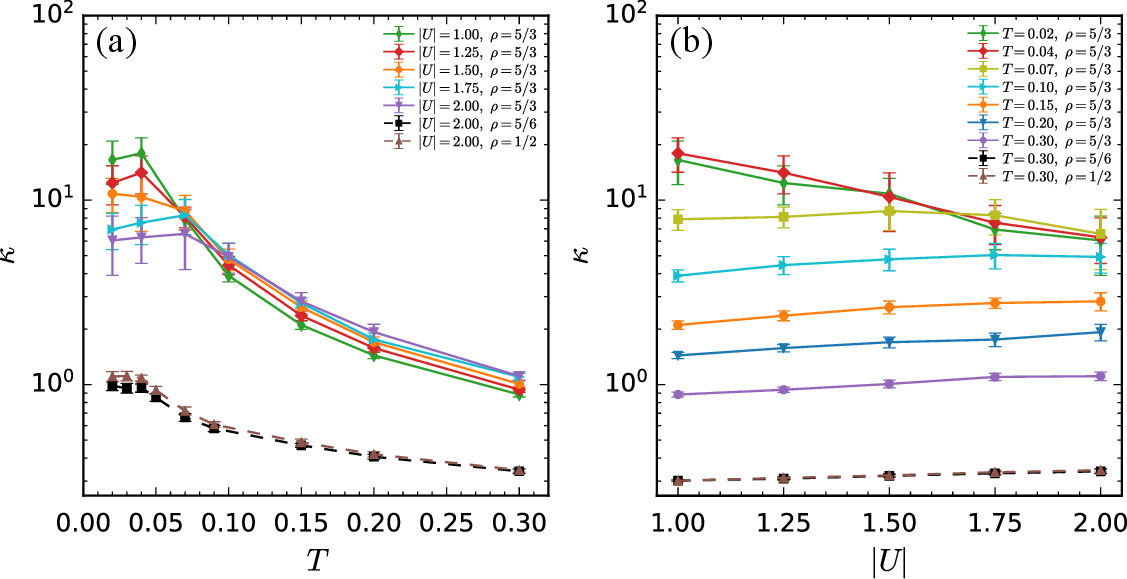}
  \caption{\label{fig:comp}Density compressibility $\kappa$ on an $L=18$ kagome lattice. (a) Temperature dependence of $\kappa$ at the flat-band filling $\rho=5/3$ for various attractive interaction strengths $|U|$ (solid lines) and at the van Hove fillings $\rho=5/6, 1/2$ for $|U|=2.00$ (dashed lines). (b) Interaction dependence of $\kappa$ at the flat-band filling for various temperatures (solid lines) and at the two van Hove fillings for $T=0.30$ (dashed lines). All lines serve as guides to the eye.}
\end{figure}
In studies of flat-band superfluidity and superconductivity, considerable attention has been given to the density compressibility~\cite{flatcomp,topoflat3,Zhangxu1},
\begin{equation}
\kappa=\frac{\beta}{N}\left[ \left\langle\left(\sum_i n_i \right)^2\right\rangle -\left\langle\sum_i n_i\right\rangle^2 \right]. 
\end{equation}
This quantity characterizes particle-number fluctuations and can be extracted from \textit{in situ} density profiles in cold-atom experiments~\cite{Tarruell2018,Cocchi2016}. We calculate the density compressibility $\kappa$ over the temperature range $0.02\leqslant T \leqslant0.3$ and the interaction regime $1.0\leqslant|U|\leqslant2.0$ on an $L=18$ kagome lattice at the flat-band filling $\rho=5/3$. For comparison, we also evaluate the temperature dependence of $\kappa$ at $|U|=2.0$, alongside its interaction dependence at $T=0.30$, for both van Hove fillings. As illustrated in Fig.~\ref{fig:comp}, the compressibility at the flat-band filling is substantially larger than that at either van Hove filling, reflecting the pronounced particle-number fluctuations inherent to the flat band.

Figure~\ref{fig:comp}(a) presents the temperature dependence of $\kappa$ at the flat-band filling $\rho=5/3$ for various interaction strengths $|U|$. For all interaction strengths considered, the compressibility $\kappa$ increases upon cooling before eventually saturating to a large value at low temperatures. This behavior is driven by pair formation and pair-density fluctuations. At intermediate and high temperatures, cooling promotes local pairing; the accompanying pair-density fluctuations generate a large density response characteristic of flat-band systems~\cite{flatcomp,topoflat3,Zhangxu1}. At low temperatures, the saturation of $\kappa$ can be attributed to an interaction-induced effective bandwidth~\cite{Zhangxu1}.

In Fig.~\ref{fig:comp}(b), we plot $\kappa$ as a function of the interaction strength $|U|$ at the flat-band filling $\rho=5/3$ for various temperatures. At intermediate and high temperatures, $\kappa$ increases with $|U|$, whereas at low temperatures, it decreases with increasing $|U|$. This crossover indicates a temperature-dependent shift in the dominant role of the interaction. While a stronger $|U|$ enhances pairing and the associated density response at high temperatures~\cite{flatcomp,topoflat3,Zhangxu1}, the large-$|U|$ suppression of $\kappa$ at low temperatures is likely caused by an interaction-induced dispersion~\cite{Zhangxu1}.

\begin{figure}[t]
  \centering
  \includegraphics[width=0.9\linewidth]{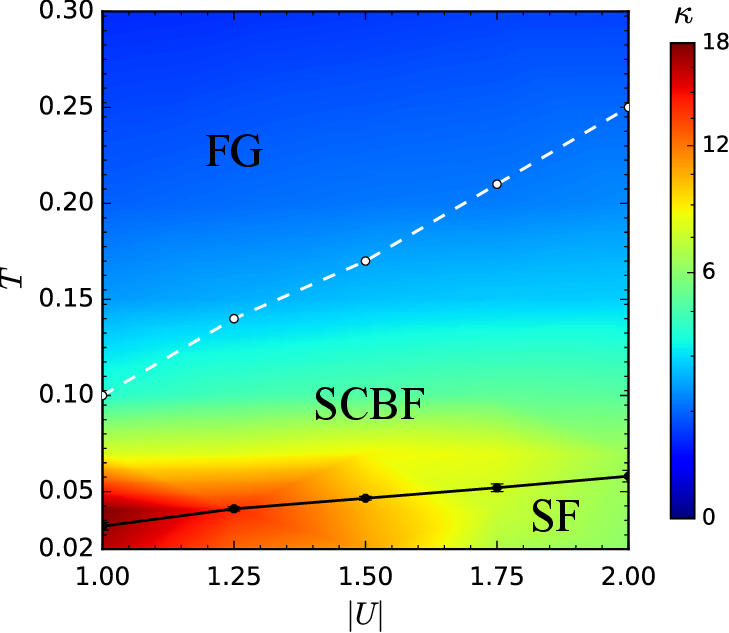}
  \caption{\label{fig:phase}Phase diagram at the flat-band filling $\rho=5/3$. The background color map indicates the interpolated density compressibility $\kappa$. The boundary of the superfluid (SF) phase $T_{\mathrm{c}}$ is marked by black data points (solid line). The crossover $T_{\mathrm{S}}$ between the normal Fermi gas (FG) and supercompressible bosonic fluid (SCBF) regions, extracted from the maximum of the spin susceptibility $\chi$, is indicated by white data points (dashed line).}
\end{figure}
\label{sec:phase}
\textit{Phase diagram.}
In Fig.~\ref{fig:phase}, we construct the phase diagram at the flat-band filling $\rho=5/3$, based on the extracted transition temperature $T_{\mathrm{c}}$, pair-formation temperature $T_{\mathrm{S}}$, and density compressibility $\kappa$. The phase diagram comprises three distinct regions separated by $T_{\mathrm{c}}$ and $T_{\mathrm{S}}$, with the background color map reflecting the interpolated values of $\kappa$. Below $T_{\mathrm{c}}$, the system resides in the superfluid phase. In the intermediate regime between $T_{\mathrm{c}}$ and $T_{\mathrm{S}}$, the spin susceptibility $\chi$ decreases upon cooling, signaling the progressive formation of spin-singlet pairs. Adopting the terminology of Ref.~\cite{Zhangxu1}, we identify this broad intermediate region as the SCBF phase, characterized by a highly compressible fluid of preformed pairs. Above $T_{\mathrm{S}}$, thermal fluctuations break the pairs, driving the system into a normal Fermi gas.

The vast separation of scales between the pair-formation temperature $T_{\mathrm{S}}$ and the superfluid transition temperature $T_{\mathrm{c}}$ originates from the quenched kinetic energy of the flat band. In the absence of single-particle dispersion, the attractive interaction dominates, driving the formation of localized bound pairs at a relatively high temperature scale $T_{\mathrm{S}}$. However, the quenched kinetic energy also minimizes the energy cost associated with changing local occupation numbers. This allows pair-density fluctuations to become exceptionally large (reflected by a drastically enhanced $\kappa$), which suppresses macroscopic phase coherence, delaying $T_{\mathrm{c}}$ to much lower temperatures. This massive mismatch between the pairing and coherence scales naturally gives rise to the broad intermediate SCBF region.


\label{sec:conclusion}
\textit{Conclusion.} We have investigated the flat-band superfluid transition in the attractive kagome Hubbard model using large-scale DQMC simulations accelerated by FFT and delay updates. By systematically evaluating the transition temperature, spin susceptibility, and density compressibility, we established the phase diagram in the weak-to-intermediate coupling regime ($1.0\leqslant|U|\leqslant2.0$) at the flat-band filling $\rho=5/3$. We find that the superfluid transition temperature scales linearly with the interaction strength, $T_{\mathrm{c}}=0.0314|U|$, a hallmark of flat-band superfluidity. Our results reveal a prominent separation of temperature scales between local pair formation ($T_{\mathrm{S}}$) and global phase coherence ($T_{\mathrm{c}}$). This separation stabilizes a broad SCBF phase between the superfluid and Fermi-gas phases, characterized by preformed spin-singlet pairs and remarkably high density compressibility. The large gap between the pair-formation scale $T_{\mathrm{S}}$ and the coherence scale $T_{\mathrm{c}}$ is fundamentally driven by the quenched kinetic energy of the flat band, which enables strong local pairing but fosters massive pair-density fluctuations (inflating the compressibility) that suppress phase coherence. A comparative analysis further demonstrates that both the transition temperature and density compressibility at the flat-band filling significantly exceed those at the van Hove fillings across the studied parameter range.
  
The steeper increase of $T_{\mathrm{c}}$ with $|U|$ compared to that in engineered topological flat-band models implies even higher transition temperatures for interactions $|U|>2$. This suggests a relatively straightforward route to flat-band superfluidity in ultracold gases: combining an intrinsic kagome flat band with an attractive Hubbard interaction. Our findings demonstrate that quantum-geometric flat bands fundamentally alter the nature of pairing and the thermodynamic response in attractive fermion systems, offering a clear physical picture of the interplay among quenched kinetic energy, preformed pairs, and phase coherence. These results provide valuable insights for ongoing experimental efforts to realize and manipulate flat-band superfluidity and superconductivity in ultracold atomic systems and kagome materials. 
As these transitions occur at parameter regimes accessible to current optical lattice experiments, this work provides a direct roadmap for observing flat-band SCBF and superfluid phases in cold-atom setups.

\textit{Acknowledgments.} This work is financially supported by the National Natural Science Foundation of China under Grant No. 12574298. We acknowledge the support of the Supercomputing Center of Wuhan University.


%
  
\clearpage
\onecolumngrid
\begin{center}
{\large\bfseries
Supplemental Material for ``Fermionic superfluid transition and pairing scale separation in the kagome flat band''}
\end{center}

\setcounter{figure}{0}
\setcounter{table}{0}
\setcounter{equation}{0}

\renewcommand{\thefigure}{S\arabic{figure}}
\renewcommand{\thetable}{S\arabic{table}}
\renewcommand{\theequation}{S\arabic{equation}}

This Supplemental Material presents supporting numerical data for the main text. Section~\ref{sec:ratio} details the finite-size scaling analysis of the dimensionless pairing correlation ratio $R_{\mathrm{Pair}}(T, L)$ for the flat-band filling $\rho = 5/3$ and van Hove fillings $\rho = 5/6$ and $1/2$ over the interaction range $1.00\leqslant |U|\leqslant 2.00$, with the extracted superfluid transition temperatures $T_c$ summarized in Table~\ref{tab:Tc}. Section~\ref{sec:error} analyzes the statistical uncertainties of $R_{\mathrm{Pair}}$ at $\rho = 5/3$ for stronger attractions $|U| = 2.5$ and $3.0$, revealing limitations that prevent reliable curve-crossing extractions and justifying the restriction of quantitative $T_c$ estimations to $1.00\leqslant |U|\leqslant 2.00$. 

\vspace{0.5cm}
\section{correlation ratio}
\label{sec:ratio}
\begin{figure}[thb]
\includegraphics[width=1.0\linewidth]{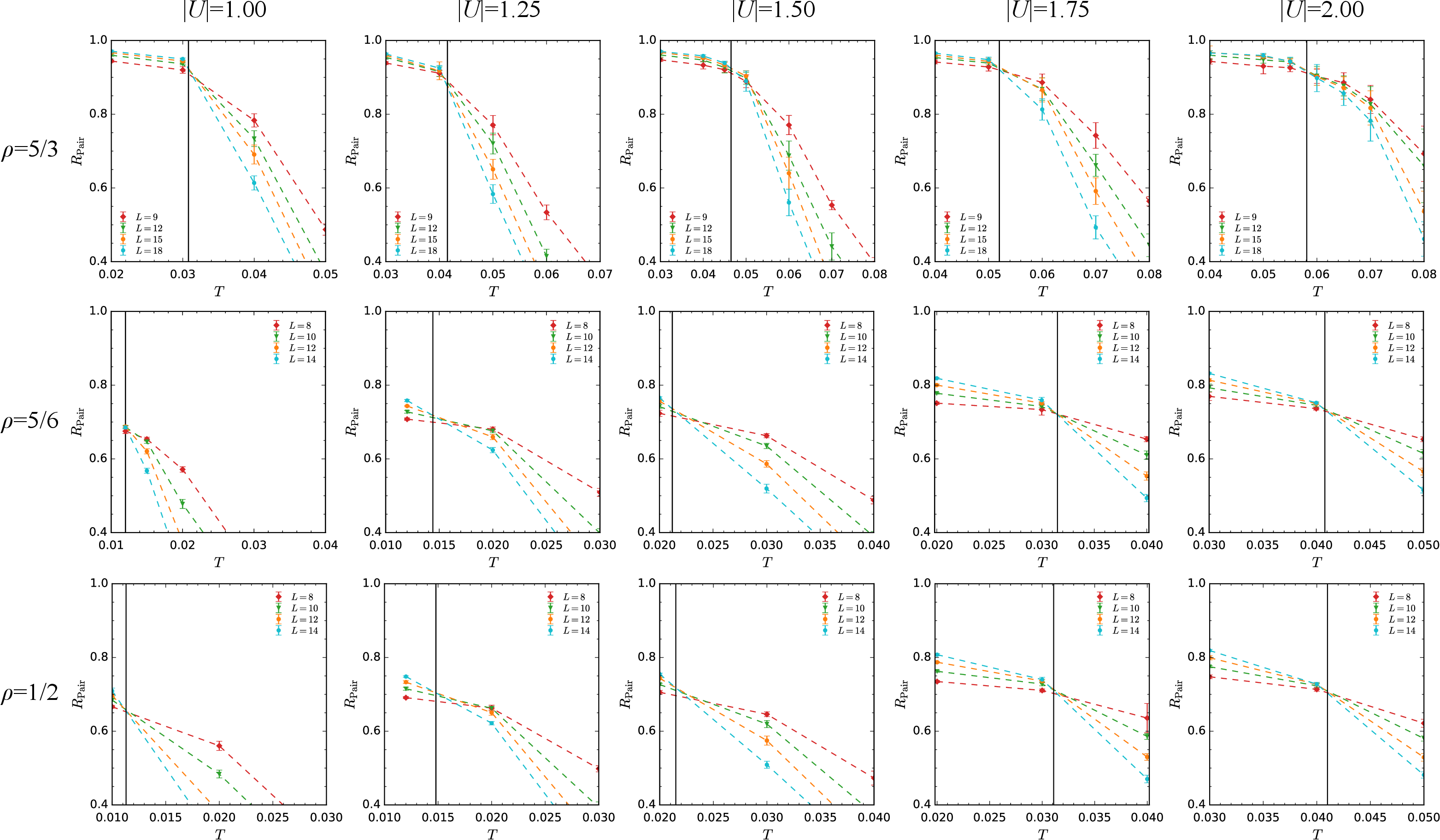}
\caption{
Temperature dependence of the dimensionless correlation ratio $R_{\mathrm{Pair}}$ for various system sizes $L$. Rows from top to bottom correspond to fillings $\rho=5/3$, $5/6$, and $1/2$, while columns from left to right correspond to interaction strengths $\vert{}U\vert{}=1.00$, $1.25$, $1.50$, $1.75$, and $2.00$. Vertical black lines indicate the estimated superfluid transition temperatures $T_{\mathrm{c}}$. Dashed lines serve as guides to the eye.
}
\label{fig:SFratio}
\end{figure}
To estimate the superfluid transition temperature $T_{\mathrm{c}}$, we analyze the finite-size scaling behavior of the dimensionless pairing correlation ratio $R_{\mathrm{Pair}}(T, L)$. Figure~\ref{fig:SFratio} shows $R_{\mathrm{Pair}}$ as a function of temperature across various system sizes $L$ at the flat-band filling $\rho=5/3$ and the van Hove fillings $\rho=5/6$ and $1/2$, for interaction strengths ranging from $|U|=1.00$ to $2.00$. Following Refs.~\cite{cross1,cross2,cross3}, $T_{\mathrm{c}}$ is estimated for each parameter set from the intersection of the curves for the two largest available system sizes. These crossing points are marked by vertical black lines in Fig.~\ref{fig:SFratio}, and the extracted numerical values are summarized in Table~\ref{tab:Tc}.
\begin{table*}[t]
\centering
\begin{tabular}{c|ccccc}
\hline\hline
$\rho \backslash |U|$
& $1.00$ & $1.25$ & $1.50$ & $1.75$ & $2.00$ \\
\hline
$5/3$
& $0.032(2)$ & $0.041(1)$ & $0.047(1)$
& $0.052(2)$ & $0.058(3)$ \\
$5/6$
& $0.0116(4)$ & $0.0144(6)$ & $0.0212(6)$
& $0.0315(9)$ & $0.0408(6)$ \\
$1/2$
& $0.0113(6)$ & $0.0148(8)$ & $0.0215(5)$
& $0.0312(7)$ & $0.0410(8)$ \\
\hline\hline
\end{tabular}
\caption{
Estimated superfluid transition temperatures $T_{\mathrm{c}}$ for various fillings $\rho$ and interaction strengths $|U|$.
}
\label{tab:Tc}
\end{table*}
\section{statistical uncertainties of the correlation ratio at flat-band filling}
\label{sec:error}
\begin{figure}[thb]
\includegraphics[width=0.5\linewidth]{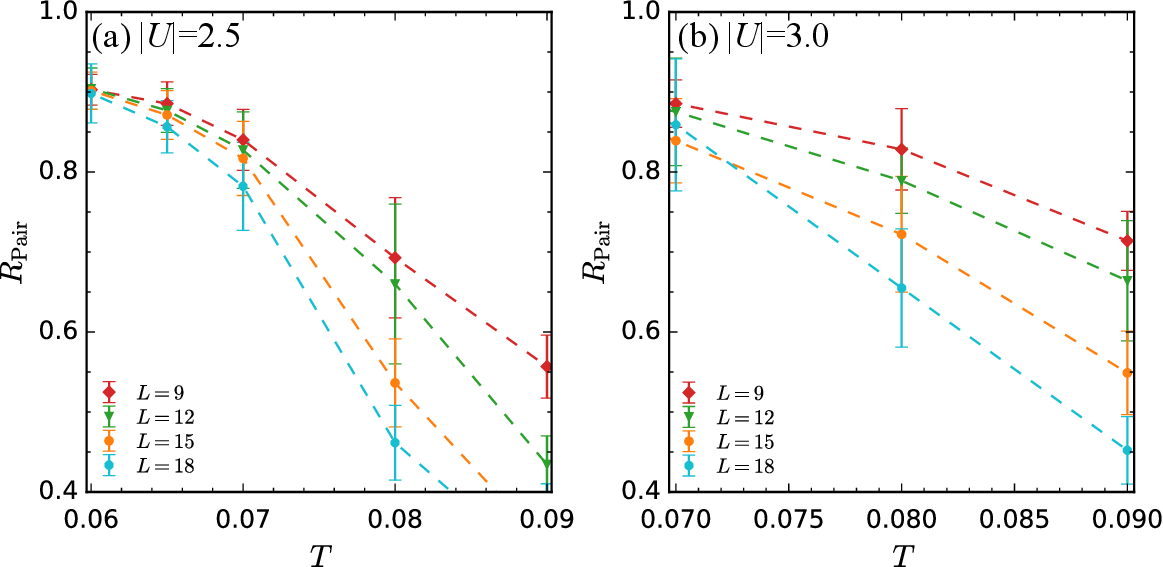}
\caption{
Temperature dependence of the dimensionless correlation ratio $R_{\mathrm{Pair}}(T, L)$ at the flat-band filling $\rho=5/3$ for system sizes $L=9$, $12$, $15$, and $18$, with interaction strengths (a) $|U|=2.5$ and (b) $\vert{}U\vert{}=3.0$. Dashed lines serve as guides to the eye.
}
\label{fig:SFerr}
\end{figure}

Figure~\ref{fig:SFerr} shows the temperature dependence of $R_{\mathrm{Pair}}(T, L)$ for system sizes $L=9, 12, 15$, and $18$ at the flat-band filling $\rho=5/3$ for stronger attractions $|U|=2.5$ and $3.0$. Large statistical uncertainties at strong couplings preclude a reliable extraction of $T_{\mathrm{c}}$ via correlation-ratio crossings. Consequently, we restrict our quantitative analysis of $T_{\mathrm{c}}$ at the flat-band filling to the interaction range $1.00\leqslant |U|\leqslant 2.00$.
\end{document}